\documentclass[twocolumn,showpacs,amsmath,amssymb,superscriptaddress]{revtex4-1}
\usepackage{graphicx}
\usepackage{hyperref}
\usepackage{breqn}
\usepackage{xcolor}
\usepackage{color}
\usepackage{booktabs}
\usepackage{float}
\usepackage{longtable}
\usepackage{epsfig}
\usepackage{bm}
\usepackage{dcolumn}

\usepackage{rotating} 

\usepackage[normalem]{ulem}
\usepackage{epstopdf}

\makeatletter
\let\cat@comma@active\@empty
\makeatother
\begin{document}

\title{Induced Monolayer Altermagnetism in MnP(S,Se)$_3$ and FeSe}

\author{Igor Mazin}
\affiliation{Department of Physics and Astronomy, George Mason University, Fairfax, VA 22030}
\affiliation{Center for Quantum Science and Engineering, George Mason University, Fairfax, VA 22030}

\author{Rafael González-Hernández}
\affiliation{Grupo de Investigación en Física Aplicada, Departamento de Física, Universidad del Norte, 081002 Barranquilla, Colombia}
\affiliation{Institut f\"ur Physik, Johannes Gutenberg Universit\"at Mainz, D-55099 Mainz, Germany}

\author{Libor \v{S}mejkal}
\affiliation{Institut f\"ur Physik, Johannes Gutenberg Universit\"at Mainz, D-55099 Mainz, Germany}
\affiliation{Institute of Physics, Czech Academy of Sciences, Cukrovarnick\'{a} 10, 162 00 Praha 6 Czech Republic}

\date{\today}

\begin{abstract}

Altermagnets (AM) are a recently discovered third class of collinear
 magnets, distinctly different from 
conventional ferromagnets (FM) and antiferromagnets (AF) \cite{Smejkal2021a,Smejkal2022x,editorial}. 
AM have been actively researched in the last few years, but two aspects so far remain unaddressed:
(1) Are there realistic 2D single-layer altermagnets? And (2) is it possible to functionalize a conventional AF into AM by external stimuli? In this paper we address both issues by demonstrating how a well-known 2D AF, MnP(S,Se)$_3$ can be functionalized into strong AM by applying out-of-plane electric field. Of particular interest is that the induced altermagnetism is of a higher even-parity wave symmetry than expected in 3D AM with similar crystal symmetries. 
We confirm our finding by first-principles calculations of the electronic structure and magnetooptical response. We also propose that recent observations of the time-reversal symmetry breaking in the famous Fe-based superconducting chalchogenides, either in monolayer form\cite{Zakeri2023} or in the surface layer\cite{MOKE,tsvelik,Josephson}, may be related not to an FM, as previously assumed, but to the induced 2D AM order. Finally, we show that monolayer FeSe can simultaneously exhibit unconventional altermagnetic time-reversal symmetry breaking and quantized spin Hall conductivity indicating possibility to research an intriquing interplay of 2D altermagnetism with topological and superconducting states within a common crystal-potential environment.

\end{abstract}

\maketitle

\section{Introduction}
One of a longest misconception in condensed matter physics, which was shattered only a few years ago, is that compensated collinear magnets cannot break time-reversal symmetry (TRS) in electronic structure. In fact, anomalous magnetic response resulting from TRS breaking, such as anomalous Hall effect or magnetooptical Kerr effect (MOKE) have been routinely used to identify ferromagnets (FM) whenewer direct measurements of magnetization are impractical (e.g., Ref. \cite{FeRh}). 
Recently, unconventional TRS breaking with a zero net magnetisation and colinear magnetic order was theoretically proposed \cite{Smejkal2020,Mazin2021}. The unconventional TRS breaking was predicted to lead to
an unconventional anomalous Hall effect\cite{Smejkal2020,Samanta2020,Mazin2021,Smejkal2022AHEReview} which was also experimentally verified\cite{Feng2022,Reichlova2020,Betancourt2021}. 

The universality of the unconventional TRS breaking was recognized by symmetry delineation of a third fundamental type of magnetism in addition to ferromagnetism and antiferromagnetism\cite{Smejkal2021a}. The third phase, altermagnetism, is sharply distinct from both conventional ferromagnetism, and antiferromagnetism  \cite{Smejkal2021a,Smejkal2022x}. Ferromagnets (and ferrimagnets) do not feature any symmetry operation that would connect opposite spin channels and thus can have a net magnetization. In conventional antiferromagnets, among the symmetry operations that map one spin sublattice onto the opposite sublattice is a lattice translation and/or inversion\cite{Smejkal2022x,Smejkal2016}. On the contrary, in altermagnets, the opposite spin densities are not connected by a simple lattice translation or inversion but are connected by a lattice rotational symmetry (for instance, a four-fold rotation, mirror or a glide plane)\cite{Smejkal2021a,jaeschkeubiergo2023supercell}. The resulting altermagnetic spin splitting of the bands changes signs in a manner determined by symmetry and breaks TRS, but the magnetisation integrates to zero over the entire Brillouin zone.

 The symmetries of altermagnetic d-wave spin channels are a clear analogy with unconventional d-wave superconductors. In the superconducting case, the {\it phase} of the order parameter usually (albeit not always, cf. nematic superconductivity) has a lower symmetry than its absolute value (the excitation gap). For example, in tetragonal superconductors, the order parameter may have a $d$-wave symmetry (as in high-T$_c$ cuprates), while the excitation gap has the full tetragonal, $i.e.$, the $s$-wave symmetry. Using the classification of Sigrist and Ueda\cite{Sigrist}, such a state has a $\Gamma_3^+$ or a $\Gamma_4^+$ representation. 
Possible types of AM can be classified analogically, revealing d, g, or i-wave alternating spin channel polarisation in electronic structure \cite{Smejkal2021a}.
 
 \begin{figure}[t]
	\centering
	\includegraphics[width=0.490 \textwidth]{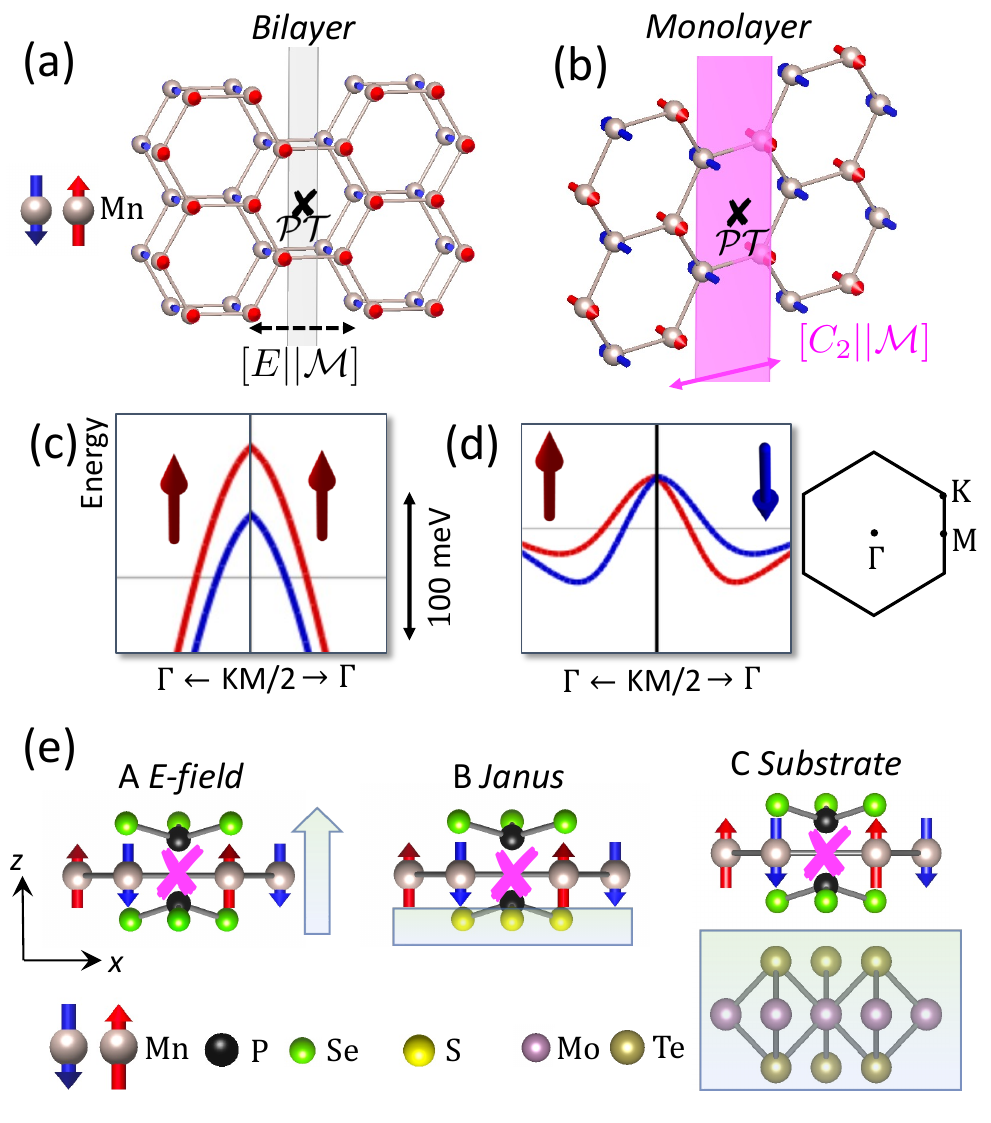} 
	\caption{Induced monolayer altermagnetism. (a) Honeycomb antiferromagnetic bilayer with marked combined spatial-inversion and time-reversal $\mathcal{PT}$ symmetry and nonrelativistic symmetry combining identity in spin space with a mirror symmetry in coordinate space $\left[E\vert\vert \mathcal{M}\right]$.
 (b) Honeycomb antiferromagnetic monolayer with marked $\mathcal{PT}$ symmetry and nonrelativistic symmetry which combines spin space rotation with a mirror symmetry in a coordinate space $\left[C_2\vert\vert \mathcal{M}\right]$. (c) Application of electric field breaks the $\mathcal{PT}$ symmetry, but preserves the mirror symmetries and results for a bilayer system in ferrimagnetic spin splitting (c) and for a monolayer system in altermagnetic spin splitting (d) in energy bands. The inset shows the labeling of Brillouin zone wavectors. $\boldsymbol{KM}/2$ marks midpoint between $\boldsymbol{K}$ and $\boldsymbol{M}$ points. (e) Three mechanism of inducing electric field asymmetry in MnPSe$_3$: A an application of external electric gate, B funcitonalizing the crystal strucure into Janus form by substituting the bottom Se layer by S atoms, and C symmetry breaking by placing the monolayer on top of a substrate. 
} 
	\label{fig1}
\end{figure}

 Curiously, altermagnetism shares some properties with AF and even more with FM\cite{Smejkal2021a,editorial,Smejkal2022,Shao2021}, while it exhibits also unique properties on their own \cite{Smejkal2020,Smejkal2021a,Ahn2019,Naka2019,Japan,Zunger,Gonzalez-Hernandez2021,Bose2022,Bhowal2022,Fernandes2023}. Remarkably, some of the unique anomalous and spin currents were experimentally already indicated\cite{Feng2022,Reichlova2020,Betancourt2021,Bose2022,Bai2021,Karube2022}. Photoemission evidence of altermagnetism was reported in circular dichroism\cite{Fedchenko2023}, direct band-splitting measurements \cite{Krempasky2023,Lee2023,Osumi2023}, and band spin splitting measurements \cite{Krempasky2023}.

The classification of AM states employed development of spin symmetries\cite{PhysRevX.12.031042}, $i.e$., pairs of symmetries in spin and coordinate space\cite{Litvin1974}, which has allowed also identification of numerous real-world materials. Notably, a great majority of those have either $d_{xy}$ or $g$-wave symmetries, respectively, 
in tetragonal and hexagonal families. The altermagnets with $d_{x^2-y^2}$ (valley polarisation\cite{Smejkal2022}) or $i$-wave altermagnets are exceedingly rare, where only a few model examples of the former have been suggested\cite{PhysRevX.12.031042,Reichlova2020,Ma2021,Smejkal2022}.
The most commonly studied altermagnetic  candidates, such as RuO$_2$, MnTe, or Mn$_5$Si$_3$, are all bulk three-dimensional centrosymmetric crystals. It is thus natural to ask whether we can identify realistic altermagnetic materials in 2D and without inversion symmetry \cite{PhysRevX.12.031042,Ma2021}.

In the present manuscript, we demonstrate a novel route towards two-dimensional altermagnetism. We show that monolayer antiferromagnets with space-time inversion symmetry $\mathcal{PT}$ can be functionalized into an altermagnetic state by, for instance, an electric field, a mechanism starkly different from the existing protocols focusing on spontaneous altermagnetic ordering in bulk crystals. 
In Fig.~1 we show that the out-of-plane electric field from gating, chemical substitution of elements ("Janusization"), or substrate influence (or surface effects), can break the $\mathcal{PT}$ symmetry while preserving altermagnetic point-group symmetries such as mirror planes connecting the opposite spin channels. 

The resulting altermagnetic bandstructure can assume the two-dimensional elusive  $i$-wave and $d_{x^2-y^2}$-wave form, as we demonstrate using real material systems: MnP(S,S\lowercase{e})$_3$ and FeSe monolayers with an altermagnetic spin splittings as large as $\sim$100 meV. 
We also show that these two-dimensional altermagnets can host strong TRS breaking response manifested $via$ large magneto-optical Kerr angles, and can explain recent experiments on monolayer and surface FeSe that indicated the presence of time-reversal symmetry breaking. Finally, we remark that as Fe-based superconductors are possible candidates for topological superconductivity\cite{Hao2014,Wang2016e,Zhang2018}, our identification of altermagnetic time-reversal symmetry breaking and spin splitting in FeSe opens a distinct unconventional route towards possible realization of topological superconductivity\cite{Zhu2023,Ghorashi2023}. Unlike the recent model proposals \cite{Zhu2023,Ghorashi2023}, here we realize the required altermagnetic time-reversal symmetry breaking intrinsically in the realistic crystal potential of superconducting monolayer FeSe.

\section{Results: M\lowercase{n}P(S,S\lowercase{e})$_3$}

\subsection{General Considerations}

 \begin{figure*}[t]
	\centering
	\includegraphics[width= \textwidth]{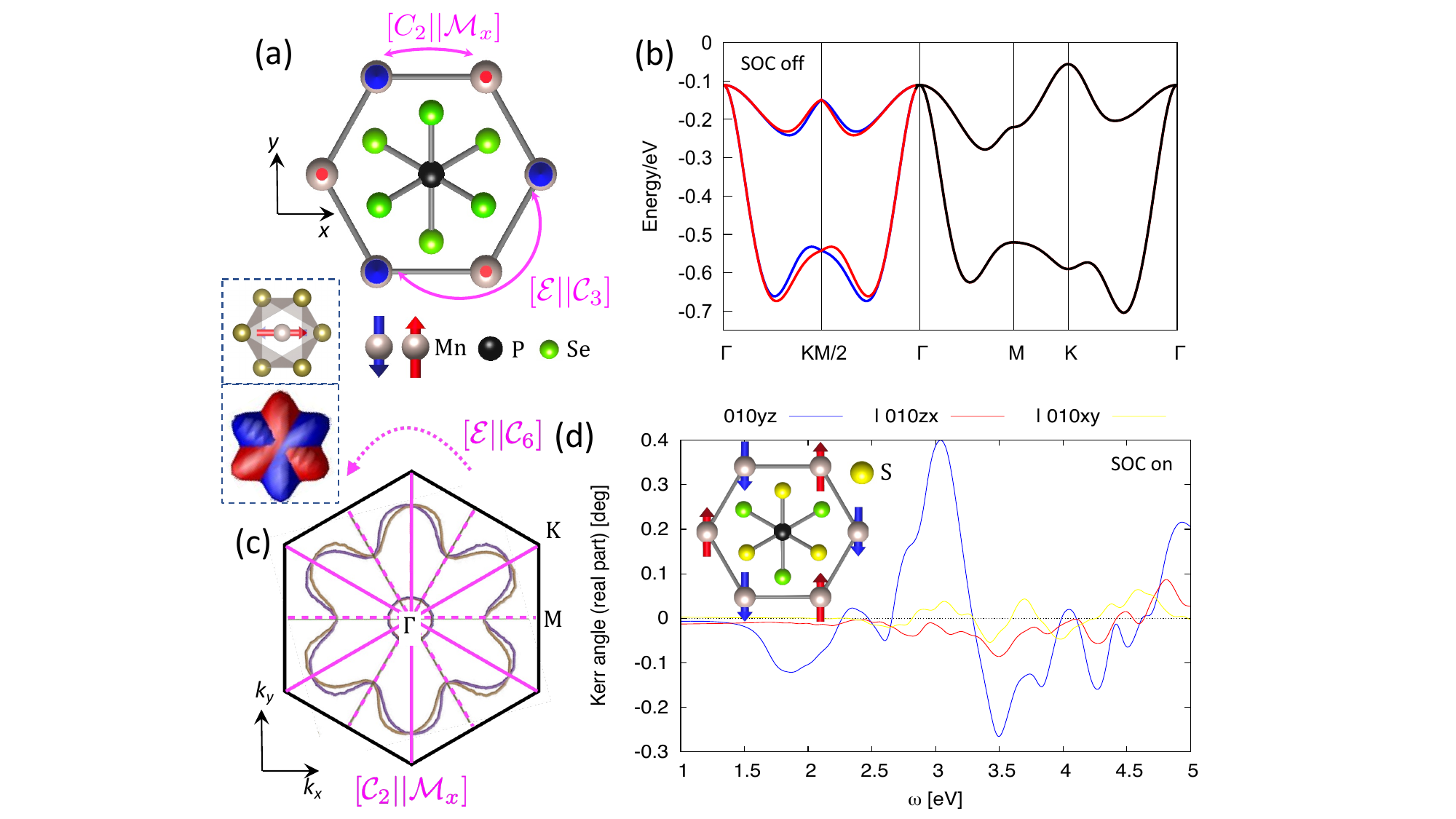} 
	\caption{Electrically induced altermagnetism in antiferromagnet MnPCh$_3$. (a) Crystal structure of MnPSe$_3$ with nonequivalent top and bottom Se layer. The system exhibits $\left[ \mathcal{C}_{2}\vert\vert\mathcal{M}_{x} \right]$ symmetry but breaks the combined space-inversion and time-reversal symmetry centered at the P atom. Calculated band structure with spin-orbit coupling switched-off shows altermagnetic spin splitting(b) of i-wave symmetry (c). (d) Calculated Kerr angle for magnetic moments along $y$ axes and spin-orbit coupling switched on. The insets in panels (a), (b) marked by dashed boxes illustrate the similar hexagonal crystal structure of MnTe with distinct altermagnetic g-wave spin-polarisation of Fermi surfaces (see details in Fig.~\ref{fig6}).}   
	\label{fig2}
\end{figure*}

MnPCh$_3$ (where Mn can be substituted by other magnetic 3d ions, and the chalcogene Ch can be Se or S or a mixture of both) is a rather unique family of van der Waals materials, easily exfoliable: they form, as opposed to most other 2D magnets, an antiferromagnetic pattern, corresponding to the standard collinear N\'eel order on the honeycomb lattice (analogues with other 3d metals may have different patterns, but here we concentrate on the Mn compound \cite{Chittari2016}).

This in-plane antiferromagnetism is different from the A-type AF found in several other 2D van der Waals compounds, such as CrI$_3$ (in a bilayer), CrSBr, MnBi$_2$Te$_4$, etc., where individual layers order ferromagnetically, but stack antiferromagnetically. In a single layer they are all FM, but in bilayers are AF --- as long as the two partner layers are symmetry-equivalent. If this symmetry is broken, for instance, by electric field gating, they become inequivalent, and thus formally ferrimagnetic (whether or not they have a nonzero net magnetization). As ferrimagnets, they break TRS, and, in particular, have Kramers degeneracy lifted everywhere in the Brillouin zone (see Fig. \ref{fig1}a,c). This effect has been observed experimentally in gated  MnBi$_2$Te$_4$ \cite{MnBiTe}. 

With this in mind, one may ask a question, what would happen if a similar experiment were performed on a single layer AF, MnPCh$_3$, where an individual plane is an AF? To answer this question, we need to take a closer look at its crystal structure (Fig. \ref{fig1}(e) and \ref{fig2}(a)). It has the space group \#162, P$\bar{3}$1m. One can see that there are multiple symmetry operations that map one spin sublattice onto the other (and many that don't): inversion centers $\mathcal{P}$ between the nearest neighbor Mn-Mn bond, and in the centers of the Mn hexagons, as well as vertical mirrors $\mathcal{M}$ perpendicular to the Mn-Mn bonds and passing through $P$'s. In addition, there are a 3-fold proper rotations $\mathcal{C}_{3}$ as indicated in Fig. \ref{fig2}(a). 

As discussed in the Introduction, the presence of spin-transposing (flipping) inversions ensures Kramers' degeneracy and excludes an AM. Intriguingly, breaking the $\mathcal{PT}$ symmetry (an inversion mapping one spin sublattice onto the other)) lifts the Kramers' degeneracy, but the two opposite-spin Mn sites remain mapped to each other through the $\mathcal{M}_x$ mirror. This is the principal difference from the previously discussed A-type bilayer case because there the two spin sublattices upon electric field are crystallographically inequivalent. In the current case the opposite spin sublattices are upon gating crystallographically locally equivalent, but globally inequivalent as they are mapped on top of each other by mirror and not by inversion or translation.  

\subsection{Practical implementations}
We will now discuss possible routes to effect the symmetry breaking discussed above. One possibility, used previously to generate a ferrimagnetic response in MnBi$_2$Te$_4$\cite{MnBiTe}, is to apply a static electric field perpendicular to the Mn plane. Such gating breaks the $\mathcal{PT}$ symmetry in two different ways: first, it shifts the two Ch and two P layers in the same direction with respect to the Mn plane. Indeed, full optimization of MnPSe$_3$ in an external electric field of 1 eV/\AA (see the Methods sections for details) resulted in about 0.11(0.13)\AA\ disparity between the heights of Se(P) atoms below and above Mn --- a sizeable difference. Second, even if the structure is not re-optimized, the addition (subtraction) of an electrostatic potential above (below) the Mn layer renders the two Se layers crystallographically inequivalent, breaking the  $\mathcal{PT}$ symmetry. Note that the crystal symmetry is then lowered from P$\bar{3}$1m to just P${3}$1m, \#157. The resulting spin splitting can be as strong as 25 meV as we show in Fig.~\ref{fig2}(b). 

Let us now turn to the angular symmetry discussed in the Introduction. The symmetry in the real space translates into the same symmetry operation present in the reciprocal space (for each spin and for even partial waves). Let us consider, for instance, a known hexagonal 3D AM, MnTe (see inset of Fig.~\ref{fig2}(a,c) and Fig.~\ref{fig5}). 
It exhibits $\left[ \mathcal{C}_2 \vert\vert 
\mathcal{C}_{6}\boldsymbol{t}\right]$
nonsymmorphic screw axis and $\left[ \mathcal{C}_2 \vert\vert 
\mathcal{M}_{z}\boldsymbol{t}\right]$ mirror plane. Therefore the
spin splitting follows the same symmetry in the reciprocal space, $i.e.,$ changes sign at the $k_z\rightarrow -k_z$ transformation, and is invariant under a threefold rotation in the $k_x,k_y$ plane. The lowest symmetry representations satisfying these condition are the  $\Gamma_3^+$, $\Gamma_4^+$ mentioned above with the bulk B-4 $g$-symmetry and characteristic angular dependencies of $k_zk_y(3k_x^2-k_y^2)$ and $k_zk_x(3k_y^2-k_x^2)$. This is indeed the AM state that is being realized in MnTe\cite{Smejkal2021a}.

The $\mathcal{PT}$ symmetry of MnPCh$_3$ gets broken if the two Ch layers are not equivalent, so one may expect that the band structure in the momentum space will only be threefold ($C_3$) symmetric. However, since in the absence of the spin-orbit coupling the electronic structure in the momentum space for a given spin has to be inversion-symmetric, $E_{{\mathbf k}\sigma}=E_{-{\mathbf k}\sigma}$, the symmetry of the system enhances to a sixfold rotation. Furthermore, in a 2D system, it cannot depend on
$k_z$, therefore the electronic structure for a given spin channel cannot follow the $g$-wave symmetry, as in the bulk case. Instead, the resulting $C_6$ symmetry,  using the terminology introduced in Ref. \cite{PhysRevX.12.031042}, generates a planar class P-6 with the angular $i$-wave symmetry of $\Gamma^+_2$, discussed in the introduction, which has 6 nodal lines in the spin splitting, as seen in Fig. \ref{fig2}b.

The two Ch layers can be also made inequivalent in another way, not requiring gating, namely ``janusization''.  Under janusization we mean synthesizing a mixed S-Se compound of the formula Mn$_2$P$_2$S$_3$Se$_3$, where all S atoms are below the Mn plane, and all Se above (or there is just a disbalance between the S/Se ratios above and below). Such Janus layers are well known in the physics of transition metal dichalcogenides\cite{JanusTMD}. In order to check the scale of the effect in this case, we have, again, optimized such a Janus structure (the space group is, of course, the same, P${3}$1m). 
In Fig.~\ref{fig7}(b) we show the calculated band structure for the  Janus case. We calculate the energy bands without SOC (the red and blue colours in Fig.~\ref{fig7}(b)) and with SOC for two different orientations of the spin quantisation axes. The magnetic symmetry for the in-plane system allows for an anomalous Hall effect and magnetooptical Kerr effect. We show in Fig.~\ref{fig2}(d) the calculated Kerr angle. The maximum value of the Kerr angle reaches 0.4 degrees, which corresponds to the order of magnitude of the Kerr angle in elemental ferromagnets with strong magnetization. In the present case, the origin of the effect is the altermagnetic time-reversal symmetry breaking caused by the combined effect of antiferromagnetism and Janusization.

Finally, we point out a third realization of the altermagnetic splitting MnPCh$_3$: engineering a heterostructure of MnPCh$_3$ with another 2D material. In fact, the lattice parameter $a=6.07$ \AA\ of MnPS$_3$ closely matches that of a well known van der Waals semiconductor, MoTe$_2$, with $a'=3.52$ \AA. Note the $a'\sqrt{3}=6.09$, matching $a$ within 0.3\%. The resulting heterostructure, shown in  Fig.~\ref{fig1}(e), has the formula Mn$_2$P$_2$S$_6$Mo$_3$Te$_6$. The calculated electronic structure (Fig.~\ref{fig7}(a)) is semiconducting, with the gap formed by  MoTe$_2$ states, but deeper lying states are derived from Mn and show sizeable spin splitting (and measurable magneto-optical response in the corresponding frequency range).

\section{F\lowercase{e}(S\lowercase{e},T\lowercase{e})}
\subsection{General Considerations}
FeSe-FeTe alloys are among the most intriguing Fe-based superconductors. Pure bulk FeSe shows no magnetic transition, but a nematic transition, often associated with stripe ($\mathbf{Q}=\pi,0$) spin fluctuations, and density functional theory (DFT) calculations result in the mean field ground state breaking the $\mathcal{C}_4$ symmetry with $\mathbf{Q}=\pi,\pi/2$ and nearly degenerates with similar states at $\mathbf{Q}=\pi,\kappa$, $0\le \kappa\le\pi/2$\cite{Jim}.
FeTe forms a unique double-stripe structure corresponding to $\mathbf{Q}=\pi/2,\pi/2$. The mixed alloys FeSe$_{x}$Te$_{1-x}$ have the highest superconducting critical temperature when $x\approx 0.5$, and there is no long-range ordered magnetism in this range.  

Arguably the most intriguing member of this family is the single-layer FeSe grown epitaxially on SrTiO$_3$\cite{FeSe-review}. It was claimed to have a critical temperature close to 100 K, when optimally doped, and is insulating and probably magnetic when undoped. Interestingly, standard DFT calculations, even with Hubbard correlation added on a static (LDA+U) or dynamic (LDA+DMFT) level, fail to reproduce the experimentally observed Fermi surface. Indeed, the latter shows rather reduced compared to DFT electron pockets around the M points in the Brillouin zone and does not have any hole pockets around the $\Gamma$ point at all.

 \begin{figure}[t]
	\centering
	\includegraphics[width=0.5 \textwidth]{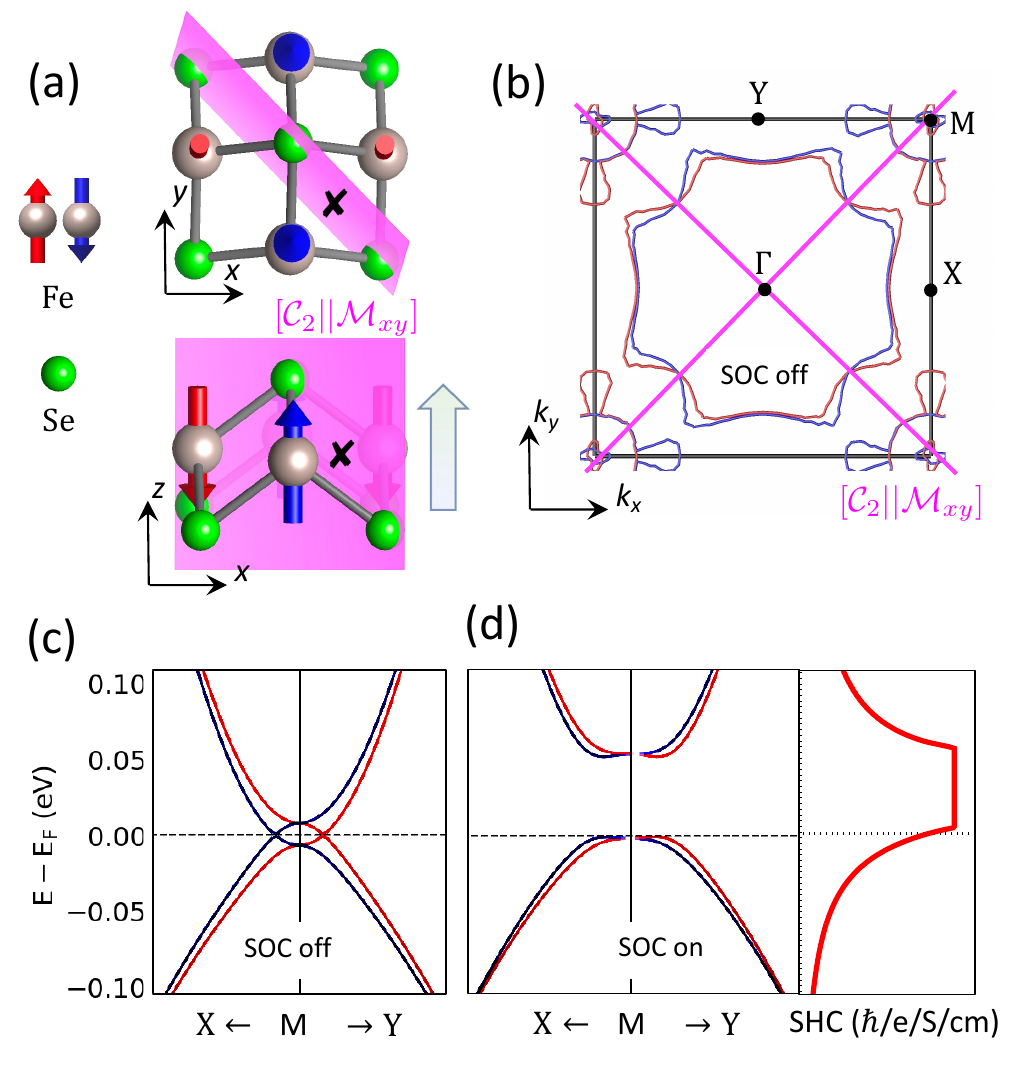} 
	\caption{Electric field induced altermagnetism in FeSe monolayer. (a) Top and side view of the crystal structure of FeSe with a checkerboard antiferromagnetic order. Application of out-of-plane electric field (marked by light green-blue arrow) violates the $\mathcal{PT}$ symmetry marked by black cross but preserves $\left[ \mathcal{C}_{2} \vert \vert \mathcal{M}_{xy} \right]$ symmetries which protects altermagnetic state. (b) Altermagnetic constant energy isosurfaces at energy = -190 meV showcase $d_{x^2-y^2}$ symmetry. (c) Altermagnetic band structure and spin splitting around $\mathbf{M}$ point. (d) (left) Band structure calculated with spin-orbit coupling. (right) Calculated spin Hall conductivity exhibits quantised plateau corresponding to topologically nontrivial band-gap at the wavevector $\mathbf{M}$ opened by the spin-orbit coupling.  } 
	\label{fig3}
\end{figure}

It was observed that imposing, on the other hand, the checkerboard order ($\mathbf{Q}=\pi,\pi$) onto the single layer of FeSe leads to a considerable reduction of the $\Gamma$ pocket, and, upon applying a static $U$, to its full disappearance\cite{Coh2015,Wang2016}. Given that in all known Fe-based superconductors LDA+U considerably worsens the agreement with the experiment in such a fundamental property as magnetic moment, this improvement of the Fermi surface may be (and likely is) fortuitous, simulating some unknown physics beyond DFT. Still, it is useful as a mechanical tool to emulate experimental fermiology. 

One important piece of information that one must keep in mind is that DFT calculation provides an approximate hierarchy of magnetic orders. While there might be quantitative corrections to the latter, on the qualitative level it must be reliable. Specifically, DFT calculations predict the ground state of all Fe-based superconductors to be single stripe, with the exception of FeTe which is correctly predicted to have double stripes. The checkerboard state is always higher in energy.  The ferromagnetic state often cannot even be stabilized, and if it can, it is much higher in energy. For instance, for a fully re-optimized FeSe monolayer, the penalty for the checkerboard structure is 26 meV/Fe, and for the FM one 68 meV/Fe, as calculated in VASP. A corollary is that while it is not unimaginable that Fe(Se,Te) in a monolayer form, or in the surface layer, has a checkerboard magnetic order (even though in DFT it is somewhat higher in energy), a ferromagnetic layer simply cannot form. We will need this corollary later.

From the point of view of magnetic symmetry, stripe (or double stripe) order in Fe(Se,Te) is not AM since the spin-mapping operation is a lattice translation. On the other hand, the checkerboard order in a single Fe(Se,Te) layer, while also not AM, has as the mapping interaction spatial inversion, as well as $\mathcal{M}_{xy}$ and 
$\mathcal{M}_{\bar{x}y}$
mirrors, see Fig.~\ref{fig3}(a). Because of the latter, breaking inversion symmetry will make the material a textbook AM. Very similar to the case of MnPCh$_3$, making the chalcogenide layers above and below Fe inequivalent breaks the inversion symmetry, converting the checkerboard AF structure into AM. 

The symmetry analysis immediately shows that it belongs to the planar $d$-wave symmetry class called P-2 in Ref. \cite{PhysRevX.12.031042}, with the basis functions  $x^2-y^2$ as confirmed in our calculations in Fig.~\ref{fig3}(b) for the electrically gated monolayer.

We will now demonstrate, that switching on spin-orbit coupling in our calculations can generate topologically nontrivial splitting at the $\mathbf{M}$-point. 
For simplicity, and because here we need only a demonstration of the principle, we take now a single  FeSe monolayer and shift Se layers so as to have the Se heights consistent with the fully optimized heterostructure with STO. In Fig.~\ref{fig3}(c) we show the altermagnetic spin splitting around $\mathbf{M}$-point calculated without spin-orbit coupling. We see clear $d_{xy}$ altermagnetic spin splitting which is consistent with our results for the electrically gated monolayer shown in Fig.~\ref{fig3}(b).

In Fig.~\ref{fig3}(d) we show the corresponding band structure calculated with spin-orbit coupling and moments along [001] direction. We observe an opening of a band gap at $\mathbf{M}$-point akin to the calculations for the undisturbed monolayer with checkerboard order.  The monolayer of checkerboard FeSe exhibits space-time symmetry and was predicted to exhibit quantized spin Hall conductivity and higher order band-topology related to the presence of $\mathcal{TC}_{4}$ symmetry\cite{Luo2022,Wang2016e}.  
 When the moments are in the altermagnetic state oriented along out-of-plane direction we preserve the $\mathcal{TC}_{4}$ symmetry and also we observe nonzero spin polarisation accumulated in the valence and conduction band only along the out-of-plane axes. Thus we expect that the altermagnetic state will preserve the nontrivial topology of spin-orbit coupling induced band-gap at the $\mathbf{M}$-point. This conjecture is confirmed by our calculation of spin Hall conductivity shown in Fig.~\ref{fig3}(d). The spin Hall conductivity exhibits a quantized plateau in the energy region corresponding to the topologically nontrivial band gap at the $\mathbf{M}$-point. We also confirm the quantum spin Hall effect by calculating nontrivial spin Chern numbers for the valence band by the Prodan method \cite{Prodan2009}.

We have thus demonstrated the possibility of combining nontrivial band-topology with unconventional time-reversal symmetry breaking in an experimentally known superconducting monolayer. We will now move on to analyse recent puzzling experiments indicating the time-reversal symmetry breaking in Fe-based monolayers.

\subsection{Experimental indication of time-reversal symmetry breaking in FeSe and Fe(Se,Te)}
Recently, several papers reported time-reversal symmetry-breaking manifestations in FeSe/SrTiO$_3$ heretostructures\cite{Zakeri2023} and in the surface layer of superconducting Fe(Se,Te) compositions\cite{Josephson,MOKE}. A natural interpretation\cite{Josephson,MOKE,tsvelik} would be in terms of surface ferromagnetism; however, as mentioned above, it is the least likely of possible magnetic orders. 

Let us take a closer look at the evidence for TRS breaking reported in these papers. First, in Ref. \cite{Zakeri2023} the disparity of the electron energy loss function (EELS) for spin-up and spin-down polarized electrons was measured, and a non-zero asymmetry was found. For instance, for the incidence energy of $\sim 2$ eV an asymmetry close to 10\% was detected in the energy range of 0--200 meV. Importantly, exactly the same asymmetry was observed in the electronic background and in two phonon peaks, which indicates that it is not due to the potential presence of asymmetric charge excitation (chiral plasmon) but to the electron-photon dipole vertex. 

 \begin{figure}[t]
	\centering
	\includegraphics[width=0.5 \textwidth]{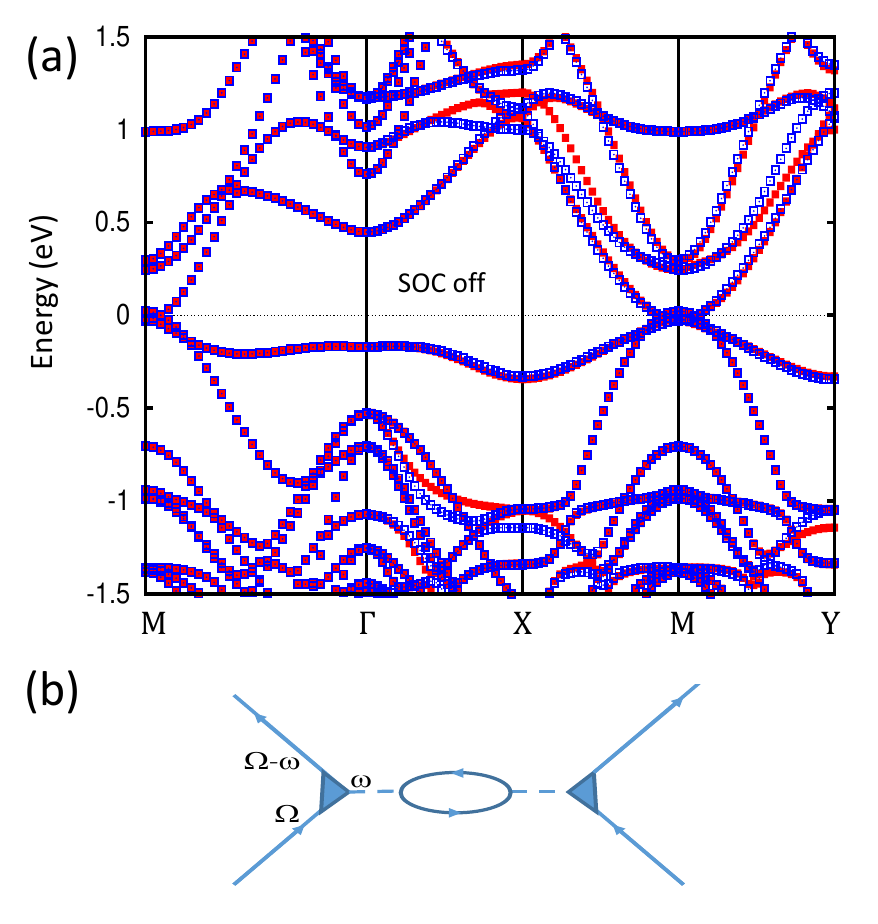} 
	\caption{Experimental manifestations of altermagnetism in monolayer FeSe on STO substrate. (a) Altermagnetic band structure calculated in a wider energy range for FeSe/STO system illustrates large altermagnetic spin splitting reaching 100 meV highlighted by the magenta rectangle. (b) Feynman diagram describing matrix elements relevant for electron energy loss spectroscopy.}
	\label{fig4}
\end{figure}

In the original paper\cite{Zakeri2023}, a model was proposed to interpret the observed chirality in a manner similar to previously known spin-asymmetry of the polarized electron diffraction (elastic scattering). Without discussing this model, we want to point out that EELS is directly related to the imaginary part of the inverse dielectric function, $-Im[1/\epsilon(\omega)]$, as shown schematically in the diagram in Fig. \ref{fig4}(b). The matrix element denoted by a triangle is by definition the same as the vortex describing the screening of electromagnetic interaction by electrons. With broken TRS and spin-orbit coupling the response may, but does not have to,  become chiral. It manifests itself in magneto-optics as well as in EELS, as different responses for left- or right-polarized light, or for spin-up and spin-down electrons. 
{The scale of these effects is set by the non-diagonal part of the dielectric function. However, for the former the relevant frequency range corresponds to the frequency at which the magnetooptical response is measured; for the latter, however, what matters is not the dielectric function at $\omega$, where $0<\omega\alt 200$ meV is the energy loss, but for the incoming electron energy $\Omega>2$ eV.}

Rather different evidence exists for the TRS breaking in the surface layer of Fe(Se,Te) with approximately 50:50 chalcogene content\cite{Josephson,MOKE}. Below we will concentrate on Ref. \cite{MOKE}, where MOKE was measured directly, and found to be small (Kerr angle on the order of 10$^{-4}$ degree), but still measurable. This TRS breaking was interpreted as ferrimagnetism in the surface layer [Fe(Se,Te) being a van der Waals material, it is highly unlikely to have a deeper level differ from the bulk in any substantial way]. It is worth noting that the penetration depth of light is $\agt 10^2$ interlayer distances, so whatever effect appears in the single surface layer it will be diluted by two orders of magnitude, or even more.

\subsection{Interpretation in terms of altermagnetism}
Here we address the question of whether van der Waals coupling to the SrTiO$_3$ substrate or the under-surface Fe(Se,Te) layer is strong enough to introduce measurable violation of Kramers degeneracy and sizeable nondiagonal response. 
This question can be answered by direct calculations which we will now discuss in connection with the experiment.

As Fig.~\ref{fig4}(a) shows, for the FeSe/SrTiO$_3$ structure there is a noticeable spin splitting $~ 100$ meV in the Fe bands, including those at high energies, relevant to EELS. Thus, the sensitivity to the electron spin in this experiment can be simply related to the AM character of the FeSe layer due to symmetry-breaking by the SrTiO$_3$ substrate. Note that it is consistent with the requirement of a sizeable SOC emphasized in Ref. \cite{Zakeri2023} (otherwise there would not be any nondiagonal component in the diagram in Fig.~\ref{fig4})(b)) and elucidates altermagnetic band splitting as the microscopic origin of the effect.

 \begin{figure}[h]
	\centering
	\includegraphics[width=0.490 \textwidth]{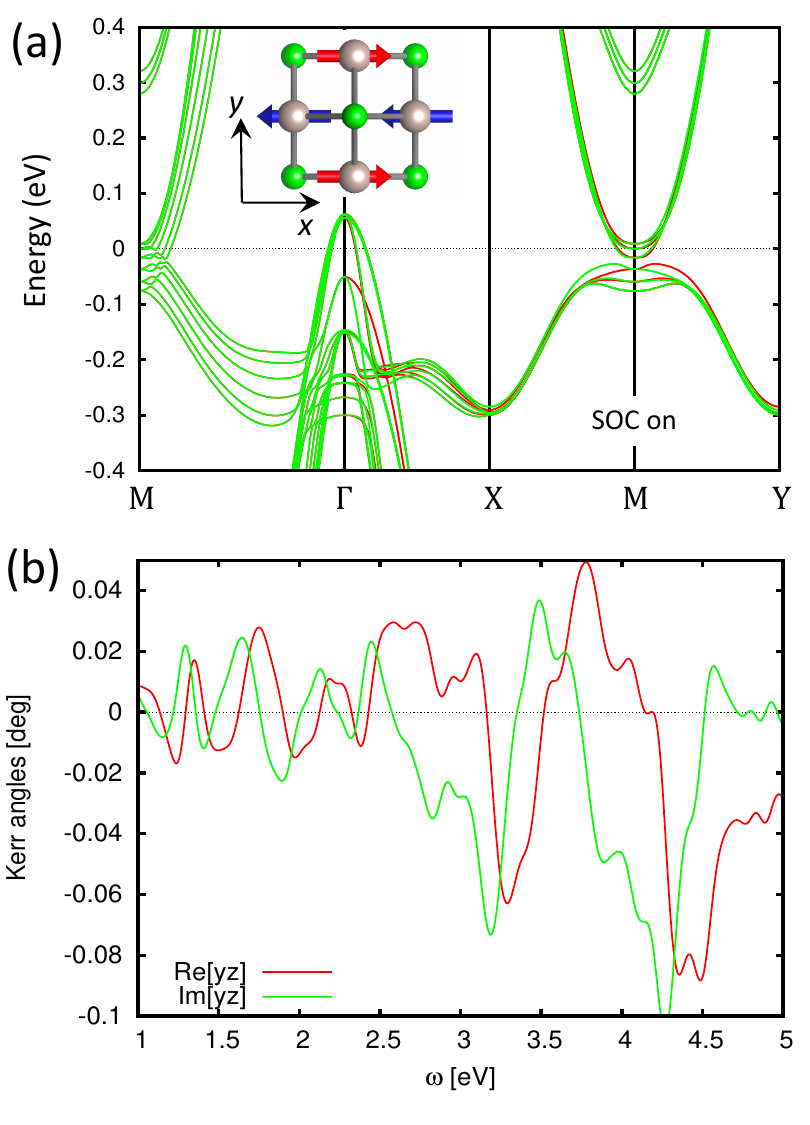} 
	\caption{Magneto-optical Kerr effect calculations for a slab of FeSe. (a) Band strucure calculated with spin-orbit coupling. The inset shows in-plane orientation of the magnetic moments. (b) Calculated magnetooptical Kerr angle.} 
	\label{fig5}
\end{figure}

Let us now consider the Fe(Se,Te) surface. Because of the Se-Te disorder straightforward DFT calculations are implausible, so we have calculated instead stoichiometric FeSe. No qualitative difference is expected with partial substitution of Se by Te. We have optimized a slab of 6 FeSe layers, consisting of two surfaces and 4 bulk layers.
In Fig.~\ref{fig5} we show the calculated energy bands of the slab with manifest altermagnetic spin splitting. In Fig.~\ref{fig5}(b) we
plot the Kerr angle calculation.  We can see a large Kerr angle on the order of 0.05--0.1 degrees, three orders of magnitude larger than in the experiment. As mentioned above, this is understandable because the penetration depth is much larger than three layers, and multiple altermagnetic domains\cite{Smejkal2020} can reduce the observed signal considerably. Thus, the observed Kerr rotation does not require a much more energetically unfavorable surface ferromagnetism but can be more naturally explained by assuming a checkerboard order in the surface layer, which leads to altermagnetic time-reversal symmetry breaking.

\section{Conclusion}
To conclude, we have presented a novel route toward altermagnetic orders in space-time symmetric monolayer antiferromagnets MnPSe$_3$ and FeSe, by various means of making the top and bottom ligand layers structurally inequivalent. The symmetry lowering breaks the space-time symmetry but can preserve the mirror symmetries that protect the altermagnetic state. We have discussed three different mechanisms: application of out-of-plane electric field, Janusization, i.e., making the top and bottom ligand layers in the structure chemically distinct, and, finally, interfacing the 2D magnet with another system that makes ligand layers crystallographically in-equivalent. 

We have also calculated magneto-optical response, which can serve as a probe of the time-reversal symmetry breaking in 2D altermagnetism. In the case of MnPSe$_3$ we predict a large Kerr angle of up to 0.4 degrees. In the case of FeSe we find that the symmetry breaking caused by the STO substrate or by the symmetry-breaking surface (including asymmetric atomic relaxation) can provide an alternative explanation for the recently reported chiral electron energy loss spectroscopy, and MOKE, respectively.

Finally, we have demonstrated that the 2d altermagnetism in FeSe is compatible with a quantized quantum spin Hall conductivity, opening  prospects for studying altermagnetism, topological states and superconductivity in two dimensions.

\section*{Methods}

Most first-principles DFT calculations were performed using 
 the Vienna ab initio simulation package (VASP) \cite{Kresse1996}, 
 with selective tests utilizing the all-electron linear augmented waves package WIEN2k\cite{Wien2k} for comparison. 
 The generalized gradient approximation (GGA) with the Perdew Burke Ernzerhof functional (PBE) \cite{Perdew1997} was utilized to account for the exchange-correlation interaction. 
 
 For Mn, we included, as typical for Mn$^{2+}$ ions, a Hubbard U term 
 in the fully-localized LDA+U limit, and in spherically symmetric approximation $U_{eff}=U-J=4$ eV. Up to 100 MnP(S,Se)$_3$ bands (correspondingly more for the heterostructure) were included in the optical calculations, and the frequencies up to 5 eV, with the K-meshes up to 26x26 in the 2D Brillouin zone. The cut-off energy was 170 eV, and the p-states of Mn were included as valence states.

 As discussed in the main text, while adding a 
a Hubbard U interaction for FeSe is more questionable, we were motivated by the need to emulate the experimental Fermiology and included in our calculations $U_{eff}=0.5 $
eV; the cutoff energy for the final calculation was 460 eV, the number of bands 240 and the optical calculations parameters were the same as in MnP(S,Se)$_3$.  
 
To study the spin transport properties, the Wannier90 code \cite{wannier90} was utilized to construct a Wannier Hamiltonian. The intrinsic spin Hall conductivity (SHC) was computed by integrating the spin Berry curvature on a dense 320x320x1 grid within the first Brillouin zone. This simulation was carried out using the \textit{linres code} \cite{linres}.

\section*{Acknowledgments}
We acknowledge fruitful discussions with Arthur Ernst and Wulf Wulfhekel. 
LS acknowledges funding from the Deutsche Forschungsgemeinschaft Grant TRR Elato-q-mat (Project A09 and B05), TopDyn funding of Johannes Gutenberg University Mainz, and the computing time granted on the supercomputer Mogon at Johannes Gutenberg University Mainz (hpc.uni-mainz.de). IM was supported by the Office of Naval Research through grant \#N00014-23-1-2480. He is also grateful to  the Wilhelm und Else Heraeus Stiftung for supporting his travel to Germany.

\newpage

\pagebreak

\section*{Supplementary Information}
 \begin{figure}[h]
	\centering
	\includegraphics[width=0.490 \textwidth]{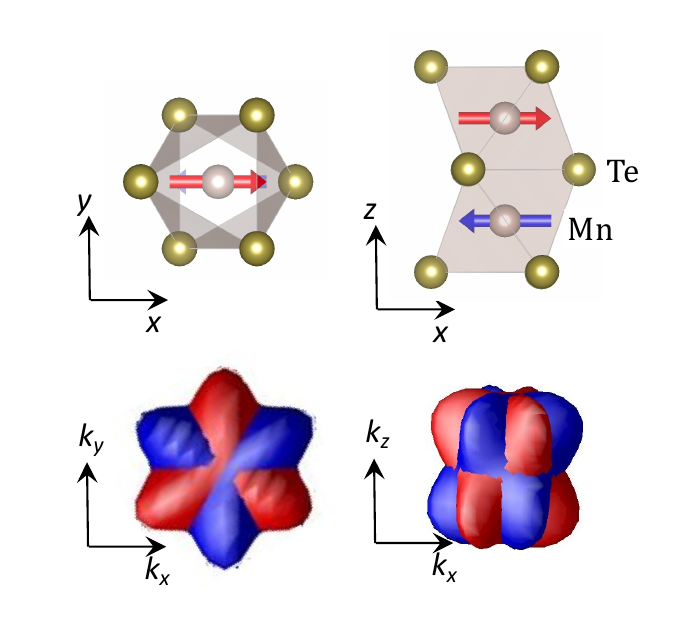} 
	\caption{Crystal structure and Fermi surfaces in MnTe.} 
	\label{fig6}
\end{figure}
 \begin{figure}[h]
	\centering
	\includegraphics[width=0.490 \textwidth]{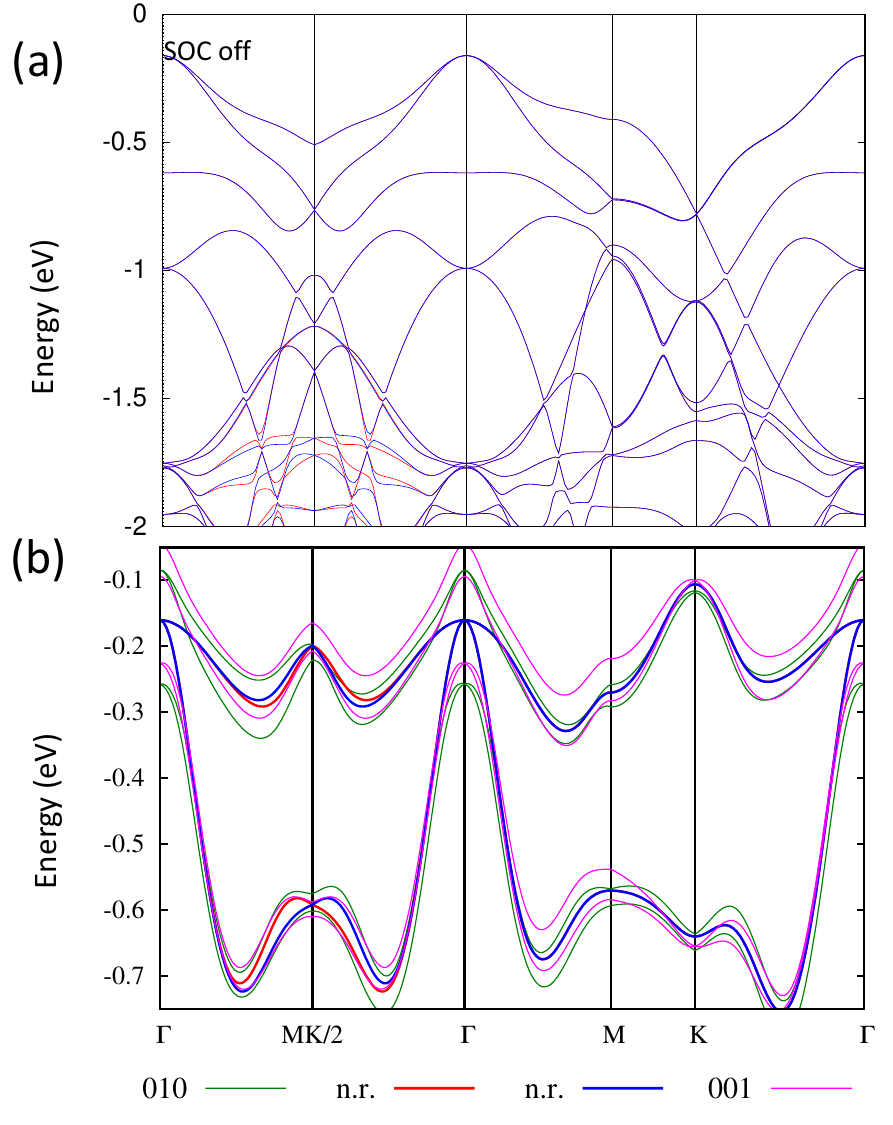} 
	\caption{(a) Band structure of MnPSe$_3$/MoTe heterostructure calculated without spin-orbit coupling. (b) Band structure of Janus Mn$2$P$2$Se$_3$S$_3$ monolayer calculated without (n.r.) and with spin-orbit coupling for the N\'eel vector along [010] and [001] direction.} 
	\label{fig7}
\end{figure}

\end{document}